\documentclass[
 reprint,
 amsmath,
 amssymb,
 aps,
]{revtex4-1}

\usepackage{graphicx}
\usepackage{dcolumn}
\usepackage{bm}

\begin{document}

\preprint{APS/123-QED}

\title{A thermodynamic approach to measuring entropy in a few-electron nanodevice.}

\author{E.~Pyurbeeva}
\author{J.A.~Mol}
\email{j.mol@qmul.ac.uk}

\affiliation{School of Physics and Astronomy, Queen Mary University of London, Mile End Road, London, E1 4NS, UK}

\date{\today}

\begin{abstract}
The entropy of a system gives a powerful insight into its microscopic degrees of freedom, however standard experimental ways of measuring entropy through heat capacity are hard to apply to nanoscale systems, as they require the measurement of increasingly small amounts of heat. Two alternative entropy measurement methods have been recently proposed for nanodevices: through charge balance measurements and transport properties. We describe a self-consistent thermodynamic framework for treating few-electron nanodevices which incorporates both existing entropy measurement methods, whilst highlighting several ongoing misconceptions. We show that both methods can be described as special cases of a more general relation and prove its applicability in systems with complex microscopic dynamics -- those with many excited states of various degeneracies.
\end{abstract}

\maketitle
\section{Introduction}
Entropy is one of the cornerstones of thermodynamics. Boltzmann's original insight in his namesake equation $S=k_{\rm{B}}\ln \Omega$ summarises the main source of power of thermodynamics -- the ability to connect a macroscopic quantity to the number of microstates $\Omega$ in a system-independent way. 

In the macroscopic realms of thermodynamics, large Hamiltonian systems with many degrees of freedom, the number of accessible microstates is so great that the microscopic meaning of entropy is largely ignored, while it is treated as a state function dependent on other, more readily measured, state functions. As the size of the system and with it the volume of its state-space are reduced, individual microstates come into focus and the knowledge of entropy can provide information about the number and relative probabilities of the microstates of the system in question. Entropy measurements have  been performed in various microscopic systems: spin-ice \cite{Ramirez1999}, 2D electron gas in GaAs structures \cite{Gornik1985, Wang1988, Bayot1996} and fractional quantum Hall states \cite{Schulze-Wischeler2007, Schmidt2017}. Yet, as experimentally accessible thermodynamic systems become progressively smaller, from quantum dots \cite{Josefssonb, Harzheim2020} and quantum dot systems, through molecules \cite{Reddy2007, Zotti2014, Cui2018} to single atoms \cite{Rossnagel2016, Lutz2020} and individual electron spins \cite{Micadei2017}, the usual approach to entropy measurement, based on the Clausius definition $dS =\delta Q/T$ becomes increasingly difficult since it involves measuring ever-decreasing heat flows. Therefore, the problem of finding an alternative entropy measurement method applicable for small systems presents itself.

Recently, two such methods were developed to measure the entropy of few-electron nanodevices. The first method relies on measuring the charge state of the nanodevice \cite{Hartman2018,Sela2019}, while the second is concerned with the electronic transport through the device \cite{Kleeorin2019, Harzheim2020, Gehring}. Here we show that both direct entropy measurement methods are special cases of a more general relation between the average electron occupation of the nanodevice and its entropy. We will derive this relation from purely thermodynamic considerations, \emph{i.e.} without any knowledge of the microscopic details of the nanodevice.

This paper is organised as follows: first, in Section \ref{system} we discuss the system under consideration and the parameters characterising it. Next, in Section \ref{rate equations} we look into the effects of degeneracy arising from the rate equation and how these have been used previously to measure entropy \cite{Hartman2018}. Then, in Section \ref{thermo1} we employ a fully thermodynamic approach to derive a thermodynamic relation for a system with no excited states and show that it describes both previously used entropy measurement methods \cite{Hartman2018, Kleeorin2019} as special cases, before expanding the approach to more complex systems with multiple excited states in Section \ref{thermo2}. Finally, we conclude with a brief summary in Section \ref{conclusion}.

\section{The system}
\label{system}
\begin{figure}
\includegraphics[width=0.5\textwidth]{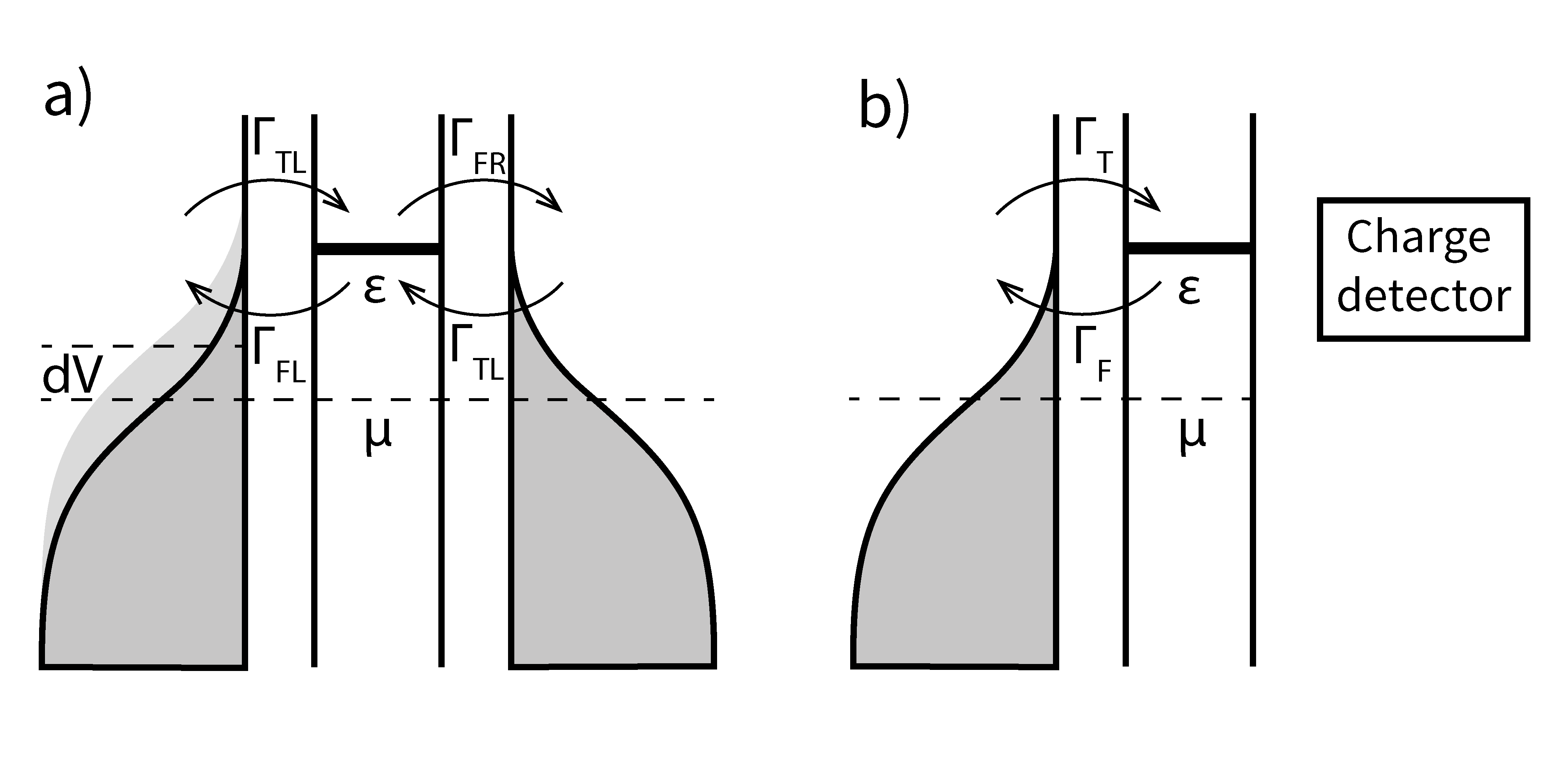}\caption{\label{fig1} Experimental regimes of Coulomb-blocked nanodevices:  a) A quantum dot coupled to a thermal bath and exchanging electrons with it. The charge state of the quantum dot can be independently determined. b) A quantum dot coupled to two electrodes through tunnel junctions. A potential difference $dV$ between can be applied between them and current through the quantum dot is measured.}
\end{figure}

Following the previously described electric entropy measurement methods \cite{Hartman2018, Kleeorin2019}, we will focus on single-electron nanodevices such as quantum dots \cite{Josefssonb, Harzheim2020} and single-molecule devices in the resonant transport regime (sequential tunnelling) \cite{Brooke2015}. Experimental measurements of these devices fall into two broad categories, as shown in Figure \ref{fig1}: charge state measurements \cite{Hartman2018}; and transport measurements (including thermoelectric transport) \cite{Kleeorin2019,Harzheim2020, Gehring}. The free parameters in both experimental setups are the temperatures of the baths and the energy level of the quantum dot (or molecule, we will refer to both as the quantum dot in the future). In the transport measurement setup (Figure \ref{fig1}b) additional degrees of freedom are the temperature difference between the baths and the bias voltage, however we will look at the quasistatic case where both are infinitesimally small.

We will consider the case where the quantum dot has only two energetically accessible charge states, with $N$ and $N+1$ electrons occupying it, and define the single-particle energy of the quantum dot $\varepsilon = E(N+1)-E(N)$ as the energy difference between the total energy of the quantum dot in the $N+1$ and $N$ charge states \cite{Nazarov2009}. The single-particle energy level can be controlled by applying a gate voltage $V_g$, $\varepsilon = \varepsilon_0 - e\alpha V_g$, where the lever arm $\alpha$ is given by the electrostatic coupling between the gate and the quantum dot. For now we will forgo the consideration of excited states and assume that energy depends on the charge state only. See Section \ref{excited} for the discussion of $\varepsilon$ in case of energy splitting of a charge state.

We treat the electrodes as ideal thermal baths with Fermi-distributions and chemical potential $\mu$. For all practical applications up to and above room temperature, the Fermi-gas in the electrodes remains highly degenerate, therefore we can put $\mu=E_F$ and neglect its dependence on temperature. Since the quantum dot and the electrodes are in equilibrium with respect to particle exchange (arbitrarily close to equilibrium in the transport measurement setup), the chemical potential of the of the quantum dot is equal to $\mu$. 

We emphasize that it does not imply that $\varepsilon$ and $\mu$ can be equated, as suggested by Hartman \emph{et al.} \cite{Hartman2018}. The single-particle energy level $\varepsilon$, the additional energy the quantum dot gets when entered by a new electron, is often referred to as an electrochemical potential (see for example \cite{Hanson2007}), however it is not one from a thermodynamic perspective. By definition, $\mu=(\partial U/ \partial N)_{S,V}$, whilst in our case, an electron entering a quantum dot necessarily changes its entropy. Below, we will demonstrate that entropy can be measured directly from $\varepsilon-\mu$. 

\section{Rate equations}
\label{rate equations}
\subsection{Degeneracy effects in the rate equation}
\begin{figure}
\includegraphics[width=0.5\textwidth]{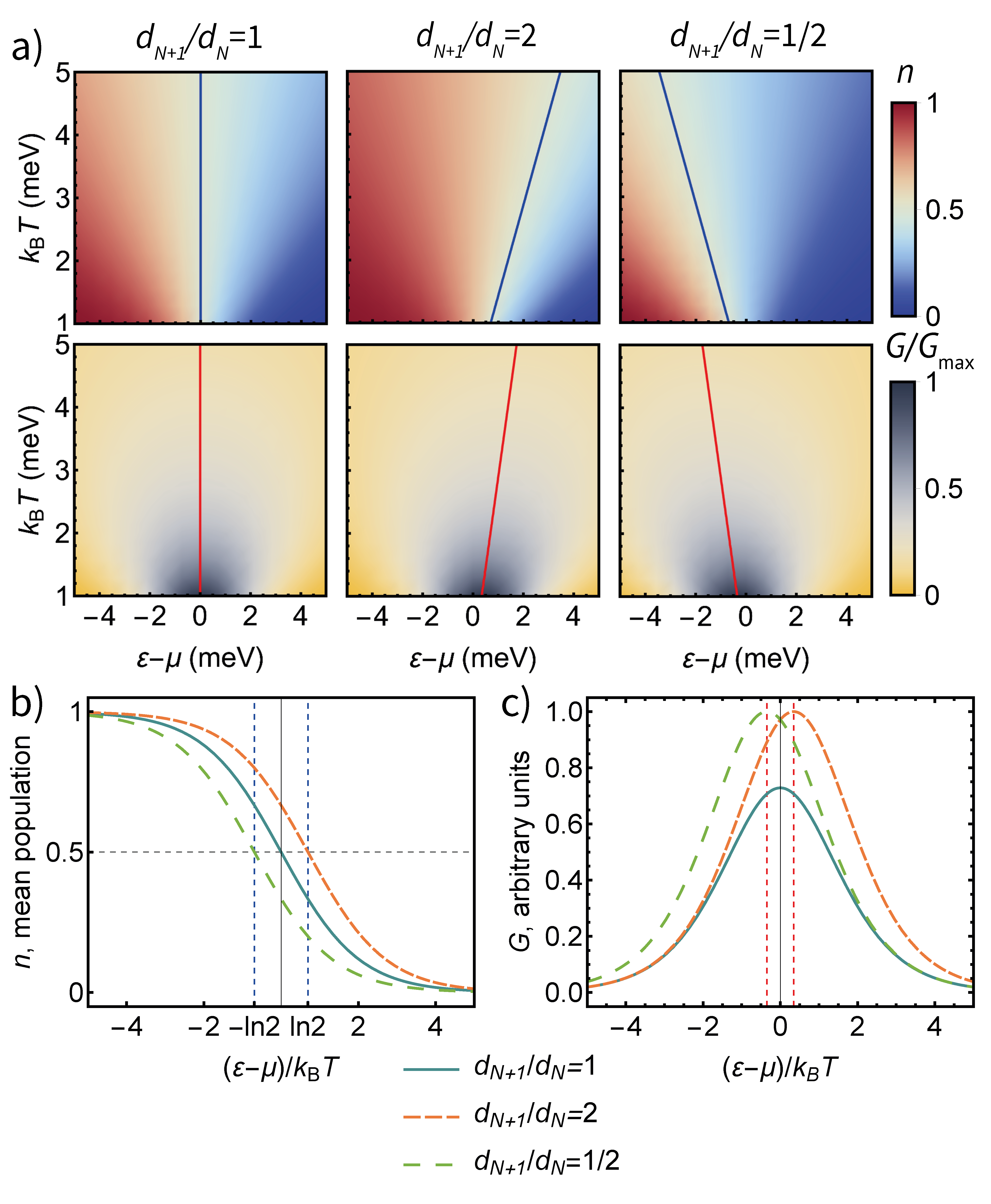}\caption{\label{fig2}a) The dependence of the mean excess population and conductance of a quantum dot coupled to a heat bath for a non-degenerate transport level and a two-fold degenerate one with even and odd $N$ respectively -- $d_{N+1}/d_{N}=$1, 2, 1/2. The blue lines show the charge degeneracy point $n=1/2$, and red lines the conductance peak for each temperature. b) The dependence of mean excess population on the dimensionless energy parameter $(\varepsilon-\mu)/k_{\rm{B}}T$ for a non-degenerate transport level and a two-fold degenerate one with even and odd $N$ respectively.  c) The dependence of the conductance of a quantum dot on the dimensionless energy parameter $(\varepsilon-\mu)/k_{\bm{B}}T$ for a non-degenerate transport level and a two-fold degenerate one with even and odd $N$ respectively.}
\end{figure}

First, we consider the effects of transport level degeneracy emergent from a rate equation approach. Starting with a quantum dot coupled to a thermal bath, with only charge states accessible containing $N$ and $N+1$ electrons, the hopping rates of the electrons to and from the dot are proportional to the degeneracies of the charge states \cite{Beckel2014}: 
\begin{equation}
\label{rates}
	\begin{cases}
		\Gamma_T=\gamma d_{N+1}f(\varepsilon)
		\\
		\Gamma_F=\gamma  d_{N}  \left[1-f(\varepsilon) \right]
	\end{cases}
\end{equation}       
where $\Gamma_{T/F}$ are the rate of electron hopping to/from the quantum dot, $f(\varepsilon)=(\exp[(\varepsilon-\mu)/k_{\rm{B}}T]+1)^{-1}$ is the Fermi-distribution of the bath, $\gamma$ is a geometric rate factor, and $d_{N/N+1}$ is the degeneracy of the charge state with $N/N+1$ electrons.

As the probability for the system to occupy a given charge state, and with it the time it spends in each state depends on the hopping rates, the probabilities $p_{N/N+1}$ of each charge state occupation can be written as:
\begin{equation}
\label{pops}
	\begin{cases}
		p_N=\dfrac{\Gamma_F}{\Gamma_T+\Gamma_F}
		\\~\\
		p_{N+1}=\dfrac{\Gamma_T}{\Gamma_T+\Gamma_F}
		\\
	\end{cases}
\end{equation} 
if the charge state of the system can be measured directly with high-enough time resolution, the fraction of the total time the system occupies a given charge state is equal to $t_N/t_{N+1}=p_N/p_{N+1}$ and contains information about relative charge state degeneracies \cite{Hofmann2016}.

In the simplest case the transport level of a quantum dot has a two-fold degeneracy due to spin orientation, the degeneracies $d_{N/N+1}$ depend on the parity of $N$ -- for an even $N$ the transport level is empty and an additional electron can have two possible spin orientations, while for an odd $N$ the transport level already contains one electron (with an arbitrary spin orientation), and an additional electron enters with an opposite spin. Therefore $d_{N+1}/d_{N}=2$ for an even $N$ and $d_{N+1}/d_{N}=1/2$ for odd $N$. The mean excess population of the quantum dot $n$ is equal $p_{N+1}$ and varies between 0 and 1. See Figure \ref{fig2}a for the dependence of $n$ on $\varepsilon$ for a non-degenerate level and two parities of $N$ of a two-fold degenerate level. 

A second charge state degeneracy effect is manifested in the quantum transport setup (Figure \ref{fig1}b). For a non-degenerate transport level, the conductance of the device is highest for the transport level coinciding with the chemical potentials of the electrodes ($\varepsilon=\mu$) -- see Figure \ref{fig2}c. The change in the hopping rates due to the level degeneracy causes a temperature-dependent shift in the peak conductance of a single-electron transistor predicted in  \cite{Beenakker1991} and experimentally measured in \cite{Harzheim2020,Gehring} (Figure \ref{fig2} a). For a two-fold degenerate transport level, standard for the spin-degeneracy of electronic current through a quantum dot  $(\varepsilon_{p}-\mu)/k_{\rm{B}}T=\pm\ln2/2$, where $\varepsilon_{p}$ is the value of $\varepsilon$ corresponding to peak conductance and its sign  depends on the parity of $N$ -- Figure \ref{fig2}c. 

Both effects -- the conductance peak shift and the charge state occupation probability depend on the degeneracies and therefore allow to construct an entropy difference between the charge states retroactively by extracting relative degeneracies. However this isn't a ``true'' entropy measurement, as it is based on assumptions about the hopping rates and in this form is only applicable to a single energy level with $d_{N/N+1}$ degeneracy in the weak coupling limit, while expansion to more complex systems, even a quantum dot in a magnetic field, is not possible, as entropy is artificially constructed utilising prior knowledge of the system.    

In order for a method to be capable of measuring the entropy difference between the charge states with arbitrary dynamics (each charge macrostate can consist of a number of microstates with different energies), and for a method to be truly thermodynamic, it has to  be free of any assumptions based on our knowledge of the system. One approach to these prior knowledge-independent, ``direct'' entropy measurements lies in applying the Maxwell relations to nanodevices.

\subsection{Detailed balance approach to Maxwell relations}
The first alternative fully-thermodynamic entropy measurement methods that did not involve the measurement of heat were developed for quantum Hall states \cite{Cooper2009, Ben-Shach2013} and utilised Maxwell relations to relate the derivative of entropy to other, more readily measurable parameters.

The idea proposed by Hartman \emph{et al.} \cite{Hartman2018} was to apply a Maxwell relation to a quantum dot device:  
\begin{equation}
\label{start}
	\left(\frac{\partial \mu}{\partial T} \right)_N=-\left(\frac{\partial S}{\partial N} \right)_T
\end{equation} 
which connects the change in entropy with the number of electrons, the quantity we are most interested in, with others, which can be measured directly, without making any previous assumptions about the nature of the system.

However, a quantum dot with few electrons is far from the usual system for application and equation \ref{start} has to be treated with utmost care.

One of the issues with the way the Maxwell relation  was used in \cite{Hartman2018} was treating the right-hand side derivative as a ratio of finite increments $\Delta S/\Delta N$, where $\Delta N=1$ since only one electron can tunnel in or out, and $\Delta S$ is the entropy change associated with a single tunnelling event. Since the quantum dot is a few-electron system and only two charge states are accessible, $\Delta N=1$ is not only a large, but the only possible fluctuation. Moreover, under the treatment of $N$ as the particle number in equation \ref{start}, its left-hand side loses its meaning, since in all states, except the two extreme ones of $\varepsilon-\mu \rightarrow \pm \infty$ the particle number fluctuates between $N$ and $N+1$ and cannot be taken as constant. 

The relationship between entropy and energy in \cite{Hartman2018} is derived from detailed balance -- if the probabilities of finding the quantum dot in both charge states are equal (the point the authors look at experimentally), the tunneling rates (equation \ref{rates}) in and out are equal, which results in the equation $d_{N+1}/d_N=(1-f(\varepsilon))/f(\varepsilon)$, which after taking a logarithm takes the form:
\begin{equation}
    \frac{\varepsilon-\mu}{T}=k_{\rm{B}}(\ln d_{N+1}-\ln d_N)=\Delta S
\end{equation}
This equation resembles the Maxwell relation written for the quantum dot, however it is only valid for one value of $\varepsilon-\mu$ -- the one corresponding to equal charge state probabilities, while a Maxwell relation holds true for all values of external parameters. 

To make sure we apply the Maxwell relation correctly to the quantum dot, in the following section we look at all the parameters involved separately.

\section{Thermodynamic relation, no excited states}
\label{thermo1}
\subsection{Derivation and entropy definition}
\label{not-excited}
First, we consider a preliminary case of a system where energy depends on the charge state of the quantum dot only -- each charge state might have several microstates, but they all have the same energy, $E(N')$, where $N'$ is equal to $N$ or $N+1$ and $E(N+1)=E(N)+\varepsilon$. 

As thermodynamics operates with averaged quantities, to derive a general thermodynamic relation between entropy difference between the two charge states of a quantum dot and its energy level, we need to consider the mean population of the quantum dot $\bar{N}$. The single-particle energy level has a mean occupation $n$ between 0 and 1, while the base population of the dot $N$ remains unchanged. Since $\bar{N}=N+n$, the mean additional energy is $\varepsilon n$ and the mean free energy $\bar{F} = E(N)+\varepsilon n - TS$. As $N$ remains constant, the derivatives in the Maxwell relation for the quantum dot can be taken by the mean excess population, which yields:
\begin{equation}
    \label{maxwell}
	\left(\frac{\partial \mu}{\partial T} \right)_n=-\left(\frac{\partial S}{\partial n} \right)_T,
\end{equation}
similarly to the ``macroscopic'' expression (equation \ref{start}). More importantly, we find the relation between the chemical potential and entropy from $\mu = \left(\partial \bar{F}/\partial \bar{N}\right)_T$, leading to:
\begin{equation}
    \label{free}
    \varepsilon-\mu = T\left(\frac{\partial S}{\partial n}\right)_T.
\end{equation}
Note that the entropy used above is the entropy of the quantum dot with a mean excess population $n$, not the entropy of one of the charge states. Next, we derive the expression this entropy. Since the quantum dot is an open system -- it can exchange both energy and particles with the environment and, in principle, an infinite number of microstates is accessible to it -- we use the Gibbs entropy expression.

In the steady state at any point of time the quantum dot exists in one of the available charge states and in one of the microstates corresponding to each of the charge macrostates. The value of entropy has to represent both macrostate and microstate uncertainty -- the uncertainty in the charge state of the quantum dot, and uncertainty in which  microstate of each charge state is occupied. 

We introduce a theorem: if a system can occupy $m$ macrostates with probabilities of occupation $p_i$ and each macrostate in turn has $m_i$ microstates with probabilities $p_{ij}$, the total Gibbs entropy of this system is 
\begin{equation}
\label{ent}
	S=S_c+ \sum\limits_i p_iS_i 
\end{equation}
where $S_c$ is the ``coarse'' Gibbs entropy of macrostate occupation $S_c=-k_B\sum\limits_i p_i \ln p_i$ and $S_i$ are the Gibbs entropies of the microstates: $S_i=-k_B\sum\limits_j p_{ij} \ln p_{ij}$.

In the case of the quantum dot with two macrostates with the probabilities $p_{N+1}=n$ and $p_{N}=1-n$ corresponding to charge states with $N+1$ and $N$ electrons, the entropy is equal to:
 \begin{eqnarray}
 \label{entropy}
 	S=-k_{\rm{B}}\left(n\ln n-(1-n)\ln(1-n)\right)+\\
 	+nS_{N+1}+(1-n)S_N. \nonumber
 \end{eqnarray}
Substituting the above, and the entropy of a  two-macrostate system into equation \ref{free}, we arrive at:
   \begin{equation}
  \label{result}
  	\frac{\varepsilon-\mu}{T}=k_{\rm{B}}\ln \frac{1-n}{n}+\Delta S
  \end{equation} 
  where $\Delta S$ is the entropy difference between the two charge states $S_{N+1}-S_{N}$ and $n$ is the mean excess population of the quantum dot. Like the initial Maxwell relation it has been derived from, this equation holds true for any value of $\varepsilon$.
  
\subsection{Applications and experimental evidence: a two-fold degenerate energy level}
 \begin{figure}
\includegraphics[width=0.5\textwidth]{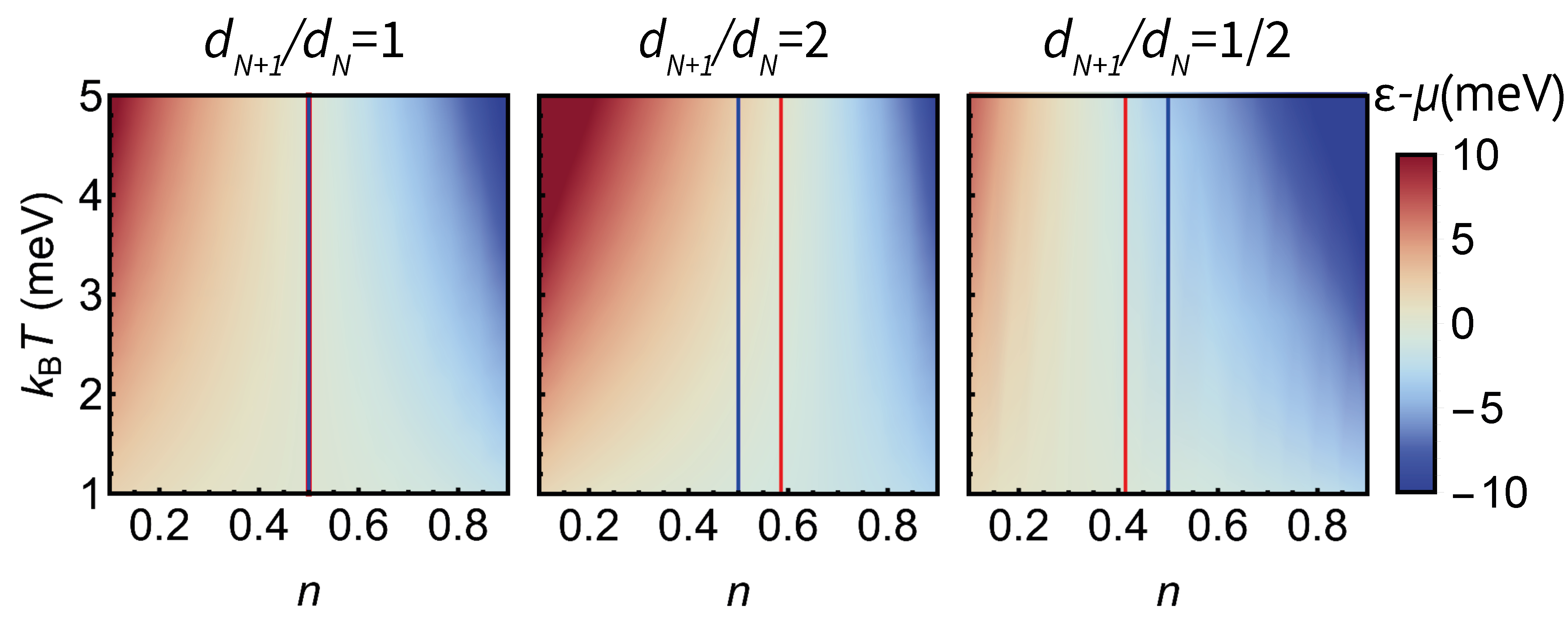}\caption{\label{fig3} Illustration of equation \ref{result} -- the dependence of the dot energy level on the population and temperature for a non-degenerate dot energy level, $d_{N+1}/d_{N}=2$ and $d_{N+1}/d_{N}=1/2$ respectively. The blue line shows the space of $\varepsilon_{1/2}$ and the red line $\varepsilon_{p}$ -- corresponding to the conductance peak. }
\end{figure}
 We have arrived at equation \ref{result} directly from the corresponding Maxwell relation without any assumptions about the properties of electronic structure in the quantum dot. It connects the entropy difference between the two charge states with the energy level of the dot, the temperature and the mean excess population of the dot as shown in Figure~\ref{fig3}.
 
 First, we will show that the general thermodynamic relation  describes the entropy measurements based on charge  and conductance. Charge state measurements \cite{Hartman2018} monitor the shift of $\varepsilon-\mu$ for the charge degeneracy point, where the probabilities of finding the system in both charge states are equal, as a function of temperature. At the charge degeneracy point $n=1/2$, the ``coarse'' entropy term $k_{\rm{B}}\ln[(1-n)/n]$ in equation \ref{result} is equal to zero, reducing the equation to $\varepsilon_{1/2}-\mu=T\Delta S$, where $\varepsilon_{1/2}$ is the value of $\varepsilon$ for the charge degeneracy point. The temperature-dependent energy for $n=1/2$ is shown in Figure~\ref{fig3} for charge state transitions where $\Delta S=0$, $k_{\text{B}}\ln2$, and $-k_{\text{B}}\ln 2$, corresponding to $d_{N+1}/d_{N}=1/1$, $1/2$, and $2/1$.
 
 Next, we show from microscopic considerations (see Appendix \ref{proof}) that the peak in conductance corresponds to the ``inverse non-degenerate'' quantum dot population: $n=1-f(\varepsilon)$ (for non-degenerate quantum dot in contact with a reservoir $n=f(\varepsilon)$). For the conductance peak, equation \ref{result} takes the form: 
\begin{equation}
  	\frac{\varepsilon_p-\mu}{T}=k_{\rm{B}}\ln \frac{f(\varepsilon_p)}{1-f(\varepsilon_p)}+\Delta S
\end{equation}
which results in $\varepsilon_p=T \Delta S/2$, agreeing with both the theoretical evaluation \cite{Kleeorin2019} for the charge transport measurement setup and the experimental result of conductance peak shifting by $\pm k_{\rm{B}}T\ln2/2$ in \cite{Harzheim2020, Gehring} for a two-fold degenerate level in a quantum dot. As shown in Figure~\ref{fig3}, for $\Delta S=0$ the conductance is maximum at $n=1/2$, while for $\Delta S = k_{\text{B}}\ln 2$ and $-k_{\text{B}}\ln2$ the population at the conductance peak is $n_p=\sqrt{2}/(1+\sqrt{2})$ and $n_p=1/(1+\sqrt{2})$ respectively. 

It is important to note that the two previously described entropy measurement methods are merely specific cases of a more general approach that allows the determination of entropy for any fixed value of the mean excess population $n$. This is particularly useful for systems where due to limited gate control not all values of $n$ are accessible \cite{Kim2014}. Moreover, equation \ref{result} can be used in reverse to find the dependence of occupation probabilities of two charge states of known dynamics (known entropy difference) on the gate voltage of the device without relying on rate equations, or when it cannot be determined, for instance when the levels and degeneracies are known, but not the hopping rates -- see appendix \ref{reverse}. 

\subsection{A single N-fold degenerate energy level}
\begin{figure}
\includegraphics[width=0.5\textwidth]{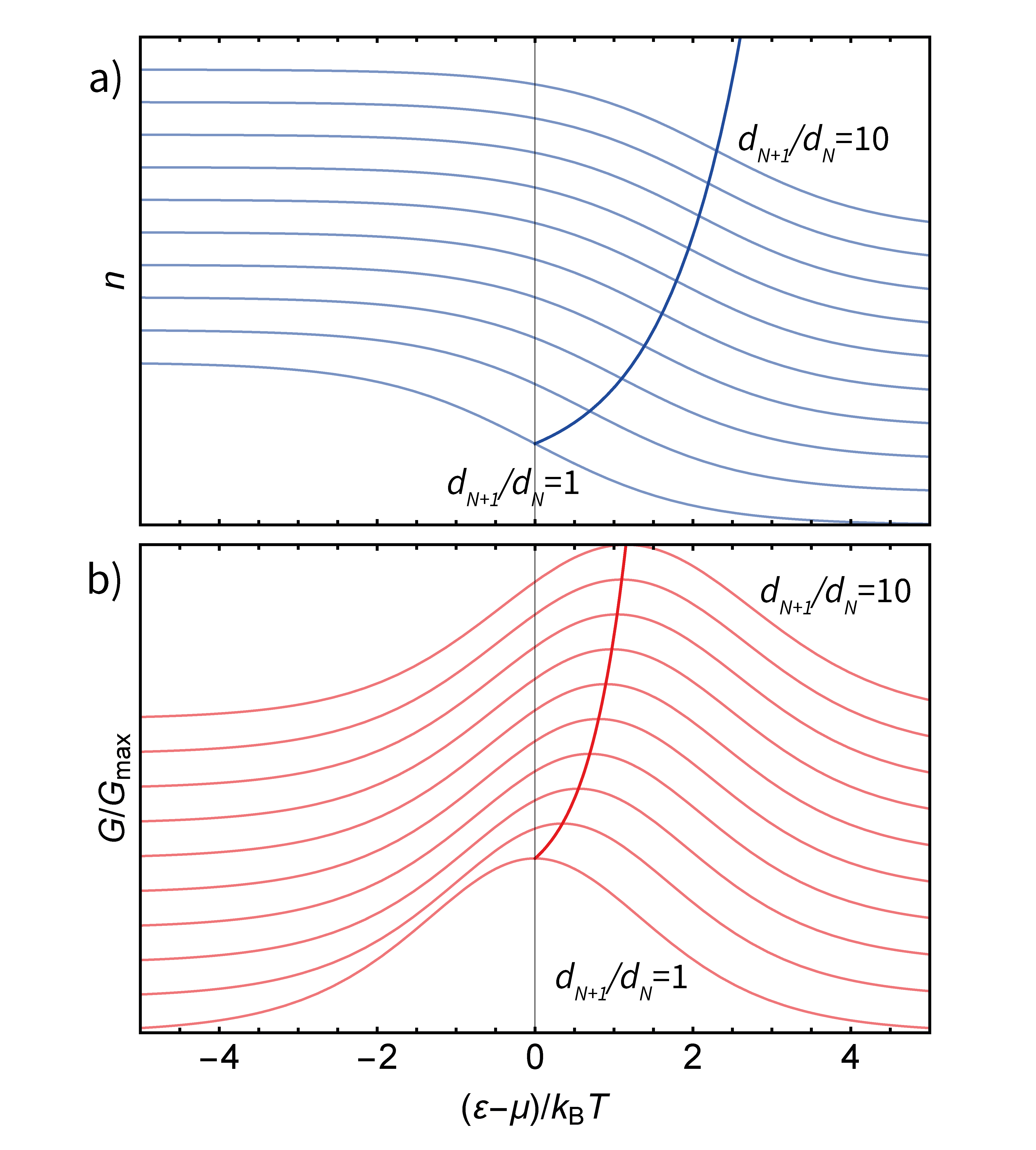}\caption{\label{fig4} a) The dependence of the population of a quantum dot on $(\varepsilon-\mu)/k_{\rm{B}}T$ for different level degeneracies. The inflection points $n=1/2$ fall on the exponential curve, as predicted. b) The dependence of the conductance of a quantum dot on $(\varepsilon-\mu)/k_{\rm{B}}T$ for different level degeneracies. The conductance peaks fall on the exponential curve with a twice greater argument.}
\end{figure}
The general thermodynamic relation can be applied to systems with a higher degeneracy of the transport level, for instance, molecules with spatial symmetry that leads to extra spacial degeneracy for each charge state. A common example of such high-symmetry molecules are fullerenes \cite{Kim2014,Sowa2019}, which have a five-fold degenerate HOMO (highest occupied molecular orbital) and a three-fold degenerate LUMO (lowest unoccupied molecular orbital).

Figure \ref{fig4} shows the dependence of population and conductance on the reduced dot level energy $(\varepsilon-\mu)/k_{\rm{B}}T$ for different values of $d_{N+1}/d_N$. For a transition between a $d_{N+1}$ degenerate state and a $d_N$ degenerate one the entropy difference is equal to $\Delta S=k_{\rm{B}}\ln d_{N+1}-k_{\rm{B}}\ln d_N=k_{\rm{B}}\ln(d_{N+1}/d_N)$. As expected, the reduced energy for the charge degeneracy point $n=1/2$ is $\Delta S/k_{\rm{B}} =\ln (d_{N+1}/d_N) $ and the conductance peak energy is $S/2k_{\rm{B}}=\ln(d_{N+1}/d_N)/2$.

\section{General thermodynamic relation}
\label{thermo2}
\subsection{Systems with excited states}
\label{excited}
\begin{figure}
\includegraphics[width=0.5\textwidth]{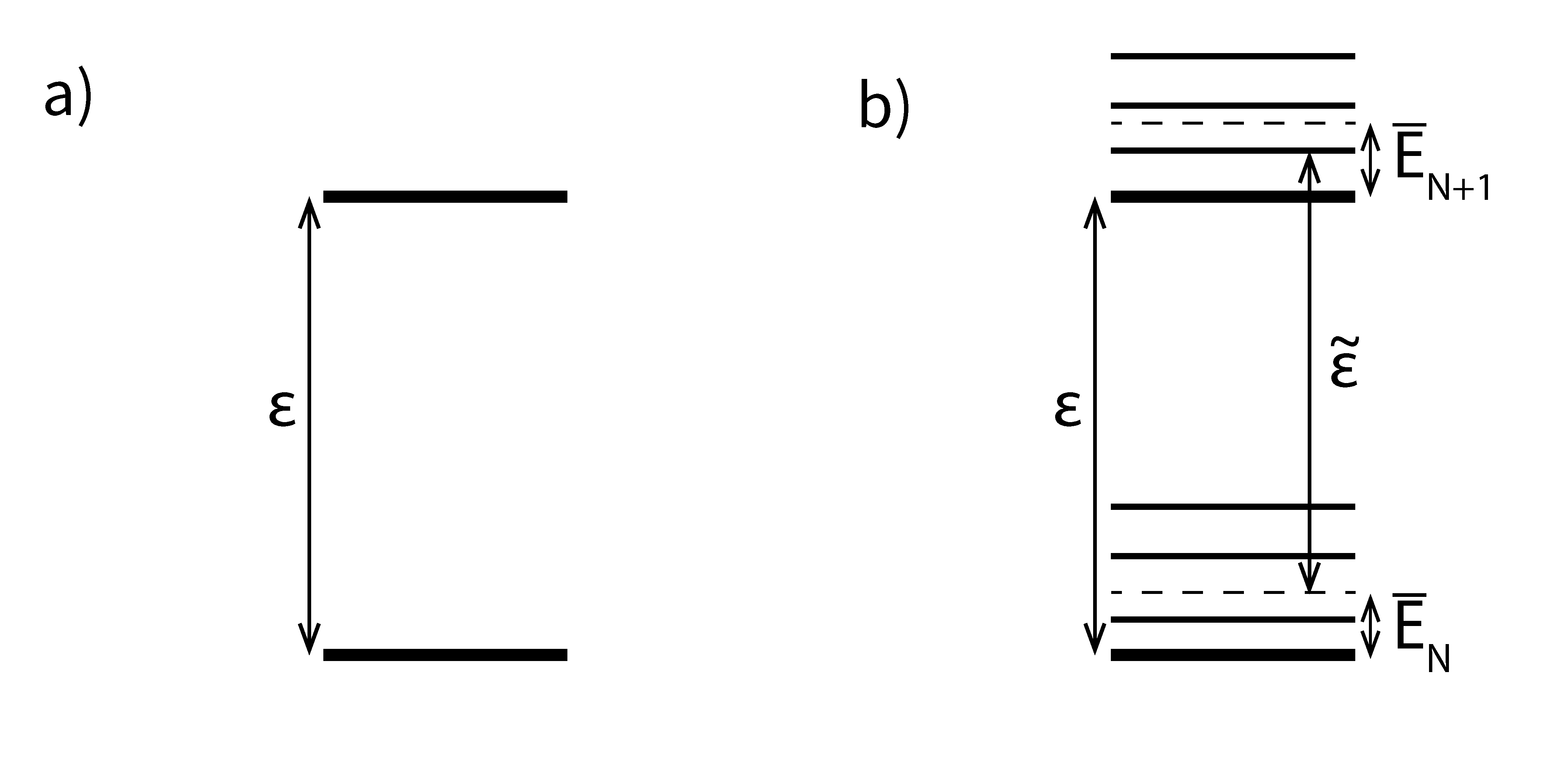}\caption{\label{fig5} a) The energy level structure of a quantum dot if all states with the same charge are energetically degenerate. b) Energy level structure in a quantum dot with excited states. Each charge state has a family of excited states at energies $E(N')+\delta \varepsilon_i$ and there is a non-zero mean energy $\bar{E}_{N'}$ of the excited states above $E_N$. $\varepsilon$ is the energy difference between the ground states, while $\tilde{\varepsilon}$ is the difference between the mean energies of the charge states. }
\end{figure}
Now we look at a more general system. Each charge state $N'$ has a family of excited states with energies $E(N)+\delta \varepsilon_i$, where $\delta \varepsilon$ can be arbitrarily large. Each of the excited states can have its own degeneracy. 

It is both usual and practical to define $\varepsilon$ as the energy difference between the ground states of the charge state families \cite{Hanson2007}. We also define the additional mean energy of the charge state:
\begin{equation}
\bar{E}_{N'}=\sum \limits_i p_i \delta \varepsilon_i
\end{equation}
where the sum is over all the states corresponding to the charge state and $p_i$ is the probability of occupation of the microstate (for the ground state $\delta \varepsilon=0$).

To write the mean free energy of the system, we need to include the mean additional energies of the charge states:
\begin{equation}
    \bar{F}=E(N)+n\varepsilon+ n\bar{E}_{N+1}+(1-n)\bar{E}_{N}-TS
\end{equation}
Following the derivation in Section \ref{not-excited}, we arrive at a new form of the thermodynamic relation:
\begin{equation}
\label{mean_energies}
    \frac{\varepsilon-\mu}{T}=k_{\rm{B}}\ln \frac{1-n}{n}+\Delta S+\frac{\bar{E}_{N}-\bar{E}_{N+1}}{T}
\end{equation}
If we define $\tilde{\varepsilon}$ as the difference between the mean energies of the charge states, $\tilde{\varepsilon}=E(N+1)-E(N)+\bar{E}_{N+1}-\bar{E}_{N}$, and equation \ref{mean_energies} can be simplified as:
\begin{equation}
\label{general}
\frac{\tilde{\varepsilon}-\mu}{T}=k_{\rm{B}}\ln \frac{1-n}{n}+\Delta S
\end{equation}
This final result could be obtained from the beginning, as the energy introduced with the additional population of the quantum dot is equal to $\tilde{\varepsilon} n$, however $\tilde{\varepsilon}$ is harder to determine, both experimentally and computationally than $\varepsilon$. We should also note that the difference between the two parameters $\varepsilon$ and $\tilde{\varepsilon}$ disappears in most experimental realisations of the measurement technique \cite{Harzheim2020,Gehring}, since $\partial \varepsilon /\partial T$ is measured.

\subsection{Discussion}
As the derivation of equation \ref{general} followed only from the Maxwell relation written for the quantum dot, it makes no assumptions about the ``nature'' of the entropy of the system -- the physical origin of the microstate probabilities and energies and the tunnelling rates into each of the microstates, and therefore it can be applied to systems with all kinds of dynamics. This is a distinguishing feature of of this work, which arises from the purely thermodynamic approach, in contrast to the previous results. 

Hartman \emph{et el.} \cite{Hartman2018} performed an experiment comparing the dependence of the thermal shift of the charge degeneracy point on the magnetic field with the theoretical expression for the entropy of a single spin in a magnetic field: $S=k_{\rm{B}}(p_{\uparrow}\ln p_{\uparrow}+p_{\downarrow}\ln p_{\downarrow})$, where $p_{\uparrow \ \downarrow}=(1+\exp(\pm g \mu_{\rm{B}} B / k_{\rm{B}}T))^{-1}$. We note that while the excellent agreement between the two is not unexpected, there has been no theoretical proof previously for applicability of the entropy measurement method for systems with transport that is not described by integer level degeneracy.
 
To further justify the thermodynamic approach, we have derived the main result of the paper (equation \ref{general}) from microscopic considerations, starting from the Gibbs distribution (see Appendix \ref{Gibbs}) for a system with excited states, to show that the ``top-down'' thermodynamic approach agrees with the more standard ``bottom-up'' microscopic one, common in the field. This also serves as evidence for the validity of our choices of entropy and chemical potential for the problem.

It may seem that the thermodynamic approach we suggest is simply a reformulation of the rate equation and, knowing the system, one can always find the shift in the charge or conductance traces. However, as the thermodynamic approach produces the value of entropy without making any prior assumptions, it can provide an important tool for choosing a physical model for an unknown system.  

One important note, however, is that while we have shown that the charge state measurement method is applicable to all conceivable systems, our proof of the validity of the method based on conductance relies on the assumption that the energy spacing between the levels corresponding to each of the charge states is small -- the hopping rates to all of them are equal (see Appendix \ref{proof}). This suggests that the applicability of the conductance measurement is narrower than that of the charge state measurement. 
 
\section{Conclusion}
\label{conclusion}
We have presented a purely thermodynamic treatment of the entropy measurement methods in few-electron nanodevices, which is free of any prior assumptions about the system. It agrees with both previously proposed entropy measurement methods, based on the charge state measurement and the conductance of the device, and, furthermore, shows that they are special cases of a single relation. This allows to broaden the experimental scope of the methods, for instance, by measuring the entropy of a system where mean population is known, but a charge-degeneracy state is not accessible. 

Additionally, we provide proof that the result holds true for much more complex systems than those that have been considered before: systems with multiple excited states with different degeneracies and large level spacings for each charge state. 

Our approach demonstrates the subtlety of applying thermodynamic relations to microscopic systems and its agreement with previous results obtained by different methods, both theoretical and experimental indicates that our application and choice of parameters is correct. Thus we can hope that we have provided a simple framework, which can be expanded for use with other microscopic systems with more than two charge states or detectable macrostates of a different origin using the same toolkit of the mean population, thermal bath-defined chemical potential and entropy that includes both microstate and charge state uncertainty.

\section*{Acknowledgements}
JAM was supported through the UKRI Future Leaders Fellowship, Grant No. MR/S032541/1, with in-kind support from the Royal Academy of Engineering.

\section*{Conflict of Interest}
The authors declare no conflict of interest.

\appendix

\section{Proof of peak conductance condition}
\label{proof}
We look at the conductance of  a quantum dot coupled to two Fermi baths (L and R). The dot has two accessible charge states, with $N$ and $N+1$ electrons, each of which can have an arbitrary number of microstates with slightly different energies within $\delta \varepsilon$ of each other. We assume that $\delta \varepsilon$ -- the degeneracy lifting is much smaller than $\varepsilon$, which allows us to treat the hopping rates to each of the charge macrostates (independent of what microstate is involved in the transfer) as dependent on $\varepsilon$ only.

The hopping rates under this assumption are:
\begin{equation}
	\begin{cases}
		\Gamma_{FR}=\gamma_F \left(1-f_R(\varepsilon)\right)\\
		\Gamma_{TR}=\gamma_T f_R(\varepsilon)\\
		\Gamma_{FL}=\gamma_F\left((1-f_L(\varepsilon)\right)\\
		\Gamma_{TL}=\gamma_T f_L(\varepsilon)
	\end{cases}
\end{equation}  
where $\Gamma_{(T/F)(R/L)}$ is the rate of electrons hopping to/from the dot with the right/left electrode involved, $f_{R/L}(\varepsilon)$ is the Fermi-distribution of the right/left electrode and $\gamma_{T/F}$ are the coefficients accounting for hopping to or from the dot, to any of the microstate levels, independent of the state of the electrodes. The occupation probabilities are the same as in equation \ref{pops}, with $\Gamma_{T/F}=\Gamma_{(T/F) L}+ \Gamma_{(T/F) R}$. 
\\
The current through the dot in a steady state is equal to 
\begin{equation}
	I=p_N \Gamma_{TL}-p_{N+1}\Gamma_{FL}=p_{N+1}\Gamma_{FR}-p_N\Gamma_{FR}
\end{equation}
We prove that the conductance peak occurs when the mean excess population of the dot is equal to the ``inverse electrode population'' $n=1-f(\varepsilon)$ by showing that the the conductance doesn't change in the first order by $ d \varepsilon $ around this point.

For $n=1-f(\varepsilon)$, the condition for the current to be zero at zero bias voltage and temperature difference is:
\begin{equation}
	\left(\frac{1-f(\varepsilon)}{f(\varepsilon)} \right)^2=\frac{\gamma_T}{\gamma_F}=e^{\frac{2(\varepsilon-\mu)}{k_{\rm{B}}T}}
\end{equation} 
Thus, $\exp[(\varepsilon-\mu)/k_{\rm{B}}T]=\sqrt{\gamma_T/\gamma_F}$ and substituting it in the Fermi-distribution, we get the occupation probabilities:
\begin{equation}
	\begin{cases}
		p_N=f(\varepsilon)=\dfrac{\sqrt{\gamma_F}}{\sqrt{\gamma_T}+\sqrt{\gamma_F}}\\~\\
		p_{N+1}=1-f(\varepsilon)=\dfrac{\sqrt{\gamma_T}}{\sqrt{\gamma_T}+\sqrt{\gamma_F}}
	\end{cases}
\end{equation}
The conductance is equal to:
\begin{eqnarray}
	G=\frac{\mathrm{d} I}{\mathrm{d} V}=\Gamma_{FR}\frac{\mathrm{d} p_{N+1}}{\mathrm{d}V}-\Gamma_{FR}\frac{\mathrm{d} p_{N}}{\mathrm{d}V}=\\~\\
	=\frac{\gamma_T\gamma_F}{2}\frac{f(1-f(\varepsilon))}{(\gamma_T-\gamma_F)f(\varepsilon)+\gamma_F} \nonumber
\end{eqnarray}
It can be demonstrated that the it does not change in the first order by $df$, which implies a zero derivative of conductance by $f(\varepsilon)$ and therefore $\varepsilon$.

\section{Dot population from the Maxwell relation}
\label{reverse}
The final expression we derive from the Maxwell relation \ref{result} can be used to find the dependence of the population of the quantum dot on  $\varepsilon-\mu$ if the entropy difference $\Delta S$ is known.

Solving equation \ref{result} for $n$ with a known $\Delta S$, we find:
\begin{equation}
\label{fermi}
	n=p_{N+1}=\frac{1}{e^{\frac{\varepsilon-T\Delta S}{k_{\rm{B}}T}+1}}
\end{equation}
This highlights the fact, first described by Beenakker \cite{Beenakker1991} that the population of the dot differs from the Fermi-distribution in the electrodes. However, the population always has the form of a Fermi-distribution shifted by some energy value and the entropy quantifies this shift.

For a two-fold degenerate level with $d_{N+1}/d_N=2$ from equation \ref{fermi} we find:
\begin{equation}
	n=\frac{1}{1+e^{\frac{\varepsilon}{k_{\rm{B}}T}+\ln2}}=\frac{2e^{-\frac{\varepsilon}{k_{\rm{B}}T}}}{2+e^{-\frac{\varepsilon}{k_{\rm{B}}T}}}
\end{equation}
which agrees with both the Gibbs distribution for the two charge states and the result found through the rate equation for arbitrary degeneracies \cite{Sowa2018a, Harzheim2020}:
\begin{equation}
\label{hard}
	p_{N+1}=\frac{d_N(\Gamma_{TL}+\Gamma_{TR})}{d_{N+1}(\Gamma_{TL}+\Gamma_{TR})+d_N(\Gamma_{FL}+\Gamma_{FR})}
\end{equation}
Comparing equations \ref{fermi} and \ref{hard} it is evident that the thermodynamic approach is \ref{fermi} yields the result in a simpler way even for a relatively simple rate equation.

 \section{Independent proof}
 \label{Gibbs}
 Our derivation of the main result of the paper, equation \ref{result} from the Maxwell relation involved several decisions on our part: mainly the use of the chemical potential and the expression for entropy. We will demonstrate that the same result can be derived directly from the Gibbs distribution, confirming the validity of our initial ansatz. 
 \\
 We have two charge macrostates, which consist of microstates with energies
 \begin{equation}
 	E_i=E_(N')+\delta \varepsilon_{i}(N')
 \end{equation}
where $N'$ can take the values of $N$ or $N+1$, and $\delta \varepsilon_i(N')$ is the energy shift of the $i$th microstate of the $N'$th charge state from the charging energy. We assume that $\varepsilon_i$ do not depend on $\varepsilon$. 

We write the Gibbs distribution for two charge states separately, taking the energy floor to be $E(N)-\mu N$, as the Gibbs distribution does not depend of the choice of zero energy:
\begin{equation}
	\begin{cases}
		P(N+1, \delta \varepsilon_i)=\dfrac{1}{Z}\Omega(N+1, \delta \varepsilon_i) e^{-\frac{\varepsilon-\mu +\delta \varepsilon_i}{k_{\rm{B}}T}}
		\\~\\
		P(N, \delta \varepsilon_i)=\dfrac{1}{Z}\Omega(N, \delta \varepsilon_i) e^{-\frac{\delta \varepsilon_i}{k_{\rm{B}}T}}
	\end{cases}
\end{equation} 
 where $\Omega(N', \delta \varepsilon_i)$ is the microstate multiplicity, and we have contracted the dependence of $\delta \varepsilon$ on $N'$ as each charge state takes its own microstate energy shifts. And $Z$ is the partition function, which has the form:
 \begin{eqnarray}
 	Z= \sum \limits_j\Omega(N+1, \delta \varepsilon_j)e^{-\frac{\varepsilon-\mu +\delta \varepsilon_j}{k_{\rm{B}}T}}+ \\
 	+\sum \limits_i\Omega(N, \delta \varepsilon_i)e^{-\frac{\delta \varepsilon_i}{k_{\rm{B}}T}}=Z_N+Z_{N+1}e^{-\frac{\varepsilon-\mu}{k_{\rm{B}}T}} \nonumber
 	 \end{eqnarray}
 where $Z_{N'}$ are the macrostate partition functions:
 \begin{equation}
 	Z_{N'}=\sum \limits_i\Omega(N', \delta \varepsilon_i)e^{-\frac{\varepsilon-\mu +\delta \varepsilon_i}{k_{\rm{B}}T}}
 \end{equation}
The mean excess population can be expressed as:
 \begin{equation}
 	\bar{n}=\frac{Z_{N+1}e^{-\frac{\varepsilon-\mu}{k_{\rm{B}}T}}}{Z_N+Z_{N+1}e^{-\frac{\varepsilon-\mu}{k_{\rm{B}}T}}}
 \end{equation}
 As $Z_N$ and $Z_{N+1}$ to not depend on $\varepsilon$ due to our prior assumption, we can solve for it, obtaining:
 \begin{equation}
 \label{res2}
 	\frac{\varepsilon-\mu}{T}=k_{\rm{B}}\ln \frac{1-n}{n}+k_{\rm{B}}\ln\frac{Z_{N+1}}{Z_{N}}
 \end{equation}
 
 In the general case, the Gibbs entropy of a system in contact with a heat bath is equal to:
 \begin{equation}
 \label{gibbs}
 	S_G=\frac{\bar{E}}{T}+k_{\rm{B}}\ln Z
 \end{equation}
 where $\bar{E}$ is the mean energy of the state. Substituting equation \ref{gibbs} into \ref{res2} we obtain the same result as we found thermodynamically (equation \ref{general}).


%

\end{document}